\documentclass[%
 aip,
 amsmath,amssymb,
 reprint,%
]{revtex4-1}

\usepackage{graphicx}
\usepackage{dcolumn}
\usepackage{bm}
\usepackage{xcolor}
\usepackage{soul}

\usepackage[utf8]{inputenc}
\usepackage[T1]{fontenc}
\usepackage{mathptmx}

\begin{document}

\preprint{AIP/123-QED}

\title[TLS Thermometer]{Sub-Kelvin Thermometer for On-Chip Measurements of Microwave Devices Utilizing Two-Level Systems in Superconducting Microresonators}

\author{J. Wheeler}
\email{Jordan.Wheeler@NIST.gov}

\author{M.R. Vissers}
 
 \affiliation{ 
National Institute of Standards and Technology, Boulder, Colorado 80305, USA
}%

\author{M. Malnou}
 \affiliation{ 
National Institute of Standards and Technology, Boulder, Colorado 80305, USA
}%
\affiliation{Department of Physics, University of Colorado, Boulder, Colorado 80309, USA}

\author{J. Hubmayr}

\author{J. N. Ullom}
 \affiliation{ 
National Institute of Standards and Technology, Boulder, Colorado 80305, USA
}%
\affiliation{Department of Physics, University of Colorado, Boulder, Colorado 80309, USA}

\author{J. Gao}

 \affiliation{ 
National Institute of Standards and Technology, Boulder, Colorado 80305, USA
}%
\affiliation{Department of Physics, University of Colorado, Boulder, Colorado 80309, USA}

\date{\today}

\begin{abstract}
We present a superconducting microresonator thermometer based on two-level systems (TLS) that is drop-in compatible with cryogenic microwave systems.   
The operational temperature range is 50-1000~mK (which may be extended to 5~mK), and the sensitivity (50-75~$\mu$K/$\sqrt{\mathrm{Hz}}$) is relatively uniform across this range.
The miniature footprint that conveniently attaches to the feedline of a cryogenic microwave device facilitates the measurement of on-chip device temperature and requires no additional thermometry wiring or readout electronics.  
We demonstrate the practical use of these TLS thermometers to investigate static and transient chip heating in a kinetic inductance traveling-wave parametric amplifier operated with a
strong pump tone.
TLS thermometry may find broad application in cryogenic microwave devices such as superconducting qubits and detectors.  

\end{abstract}

\maketitle


Cryogenic microwave techniques enable advances in a growing list of research areas.  
Examples include superconducting qubits for quantum computing\citep{Kjaergaard2020}, 
microwave kinetic inductance detectors (MKIDs) for astrophysics\citep{Zmuidzinas2012}, 
microwave readout of large arrays of low temperature detectors \citep{Mates2011} for a broad range of applications, 
and the development of quantum-limited amplifiers \citep{castellanos2008amplification,bergeal2010phase,Eom2012,Vissers2016}.
In these examples, sub-kelvin operation is essential and parasitic heating is a concern.  
Heating can originate from the device itself or from the required electrical and mechanical connections to room temperature electronics.   
For many applications, knowledge of the on-chip device temperature is important, as this quantity may differ from the heat-sink temperature for a number of reasons.  
Conventional low-temperature thermometers mounted to the device package determine heat-sink temperatures with high accuracy but are difficult to implement for on-chip temperature measurement.   
In this letter, we present a thermometer to solve the problem of cryogenic on-chip temperature measurement of microwave devices. 
In particular, this project was developed to allow for quantitative characterization of the heating in our kinetic inductance traveling-wave parametric amplifiers (KITs) \citep{Malnou2020}.

\begin{figure}[h!]
\includegraphics[trim=0.0in 0 0 0,clip,scale=.215]{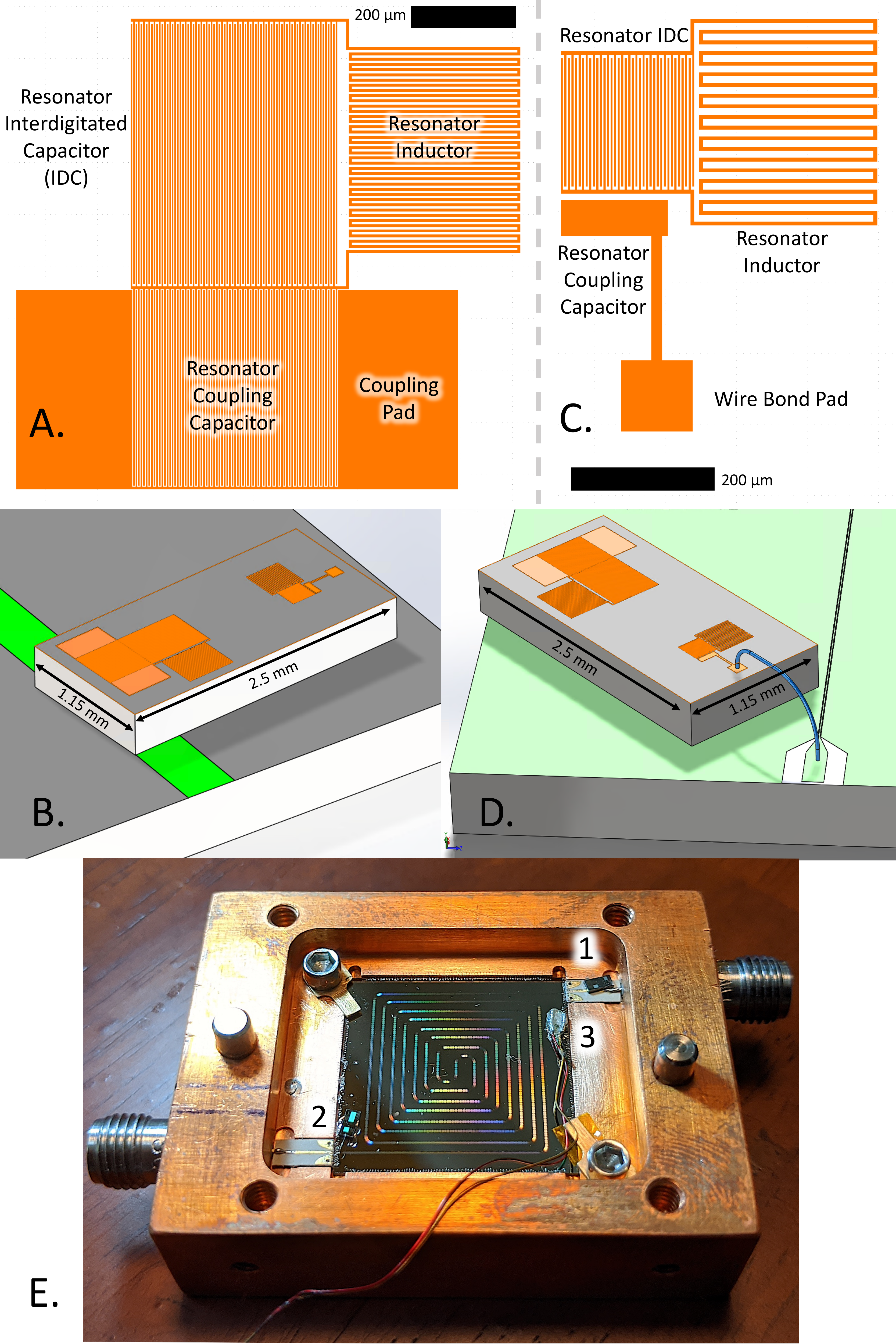}
\caption{\label{fig:design} Design and implementation of TLS thermometers.
In the type-I design (A and B), the thermometer chip is placed directly over the devices under test (DUT), and the coupling pad directly couples to the fringing field of the DUT microwave feedline.
For coupling to a DUT that lacks a significant fringing field, 
the type-II design (C and D) includes a wire bond pad that is wire-bonded to the DUT feedline end-launch.
Panel E shows the KIT package used in this work, equipped with a type-I, a type-II, and a RuOx resistive thermometer at locations 1,2, and 3, respectively.  
The RuOx and type-II thermometers are mounted on top of the KIT, whereas the type-I thermometer is placed on top of the SMA-to-device transition circuit board.
All thermometers are adhered with rubber cement.}
\end{figure}
\begin{figure*}[]
\includegraphics[scale=.45]{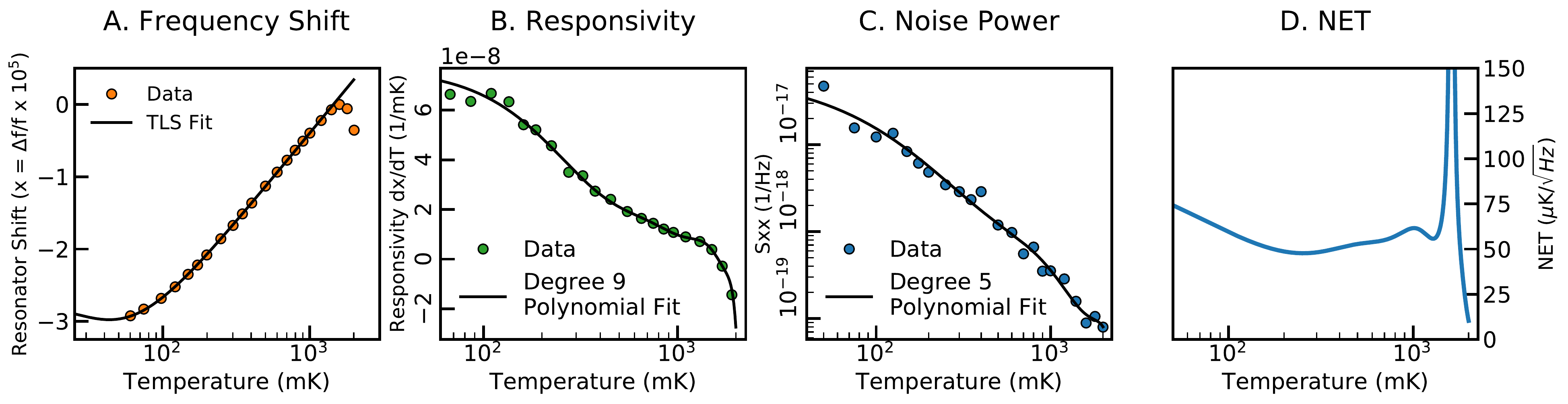}
\caption{\label{fig:sensitivity} Characterization of type-II TLS thermometer.
Panel A shows the resonator fractional frequency shift as a function of temperature and the fit to Eqn.~\ref{eqn:TLS} for T$<$1000~mK. 
The thermometer is monotonic between 50~mK and 1400~mK, and the temperature residuals of the fit are $\leq$5~mK for T$<$1000~mK. 
The fit yields $f_0$ = 2.0394 GHz $\pm$ 110 Hz and F$\delta_{TLS}^0$ = 3.37$\times10^{-5}\pm$ 1.3$\times10^{-7}$ (95\% CI).}
The slope of this shift, the responsivity, is plotted in panel B. 
Panel C shows the noise power in fractional frequency shift units, acquired with -90~dBm readout power (at which self heating is insignificant), measured at 1 Hz as a function of temperature. 
Panel D presents the noise-equivalent temperature, NET, of the resonator calculated as NET = $\sqrt{S_{xx}}$/R.
\end{figure*}

Two-level systems (TLS) are universal in lithographed superconducting microresonators and form the basis for the TLS thermometer.
Although the microscopic nature of TLS is not well understood \citep{Muller2019}, evidence suggests that one or perhaps a group of atoms in an amorphous (glassy) dielectric tunnel between two sites, giving rise to a broad spectrum of quantum tunneling states \citep{Phillips1972,Anderson1972}. 
TLS are observed in deposited dielectric layers (such as SiO$_2$ and SiN$_x$), the surface oxide of a superconductor, and the superconductor/substrate interface layer\cite{Ga02008a}. 
Moreover, the interaction between TLS and microwave photons causes resonator frequency shift, loss, and noise\cite{Zmuidzinas2012,Gaothesis}, as well as qubit decoherence \citep{Martinis2005}.

Glassy material exhibiting large temperature-dependent dielectric constants has been proposed for cryogenic thermometry \citep{Lawless1971}. Additionally, other commercial capacitive thermometers currently exist. Different from these previous implementations, we incorporated the TLS into a high Q superconducting resonator and read it out using microwave techniques.
The change in resonance frequency sourced by TLS as a function of temperature, $T$, is\citep{Gao2007,Gaothesis}
\begin{equation}\label{eqn:TLS}
\frac{f(T)-f_0}{f_0} =   \frac{F \delta^0_{\mathrm{TLS}}}{\pi}\left[\text{Re} \Psi\left(\frac{1}{2}-\frac{\hbar  f_0}{j k_B T}\right)-\text{log}\frac{\hbar f_0}{  k_B T} \right].
\end{equation}
The free parameters in this equation are $f_0$ (the resonator frequency at zero temperature) and F$\delta^0_{\mathrm{TLS}}$ (the filling factor of the TLS volume times the TLS loss tangent).
$\Psi$ is the complex digamma function.  
The TLS thermometer uses Eqn.~\ref{eqn:TLS} to map a frequency measured with a vector network analyzer (VNA) or a homodyne readout system \citep{Gao2007} to a physical temperature.  
The frequency shift, as opposed to the change in resonator quality factor, is independent of microwave power \citep{Gaothesis}. 
This allows a large range of potential readout powers.  
The frequency shift is monotonic at temperatures $T>hf_0/2k_B$ and $\leq~T_c$/8, where the thermal quasiparticles in a superconductor of transition temperature $T_c$ dominate the frequency shift \cite{Kumar2008}. 
Thus, for the $\sim$2~GHz niobium resonators demonstrated in this letter, the operating temperature range is 50--1000~mK.
Lower temperature operation may be accessed by use of resonators with $f_o < 2$~GHz.    


We have made two types of TLS thermometers, and the designs are presented in Fig.~\ref{fig:design}.
The resonators are fabricated from a single layer of 200~nm thick niobium on top of a 380~$\mu$m thick silicon wafer.  
A 25~nm layer of thermal SiO$_2$ on top of the substrate serves both as an etch-stop for the Nb SF$_6$ etch and a responsivity boost by increasing the amount of TLS.
The two thermometer types were optimized using 2D electro-magnetic simulations to achieve strong coupling to the host device feedline whose temperature is to be measured. 
Depending on the strength of the fringing field of the host device feedline, the TLS thermometer can couple directly (type-I) or indirectly via a wire bond (type-II) to the host, as described in detail in Fig.~\ref{fig:design}. 
The two types of TLS thermometers are otherwise equivalent.
Both coupling schemes require no additional wiring beyond the existing microwave feedlines and thus do not introduce any sources of electromagnetic interference (EMI) or conductive heating. 
The lumped-element resonator consists of an interdigitated capacitor (IDC) with 2.5 $\mu$m wide fingers and gaps and a 5 $\mu$m wide meandering inductor. The type-I (type-II) design has a capacitance of $\sim$2.2 pF (1.8~pF) and inductance of $\sim$5 nH ($\sim$3.5 nH).
Coupling quality factors to the DUT of 35,000 to 50,000 were obtained for the fabricated devices. 
This results in deep, easily identified, resonant features given the internal quality factors of $\sim$150,000.
Resonance frequencies ranging from 1.5 to 6 GHz were successfully fabricated.
Simulations show that the next resonances occur at $\sim$5x their base resonance frequency and the type-II version's wire bond (1~mm long) may also introduce a parasitic resonance at $\sim$15 GHz. 
The resonances are out of the band of interest of our demonstration experiment (4-8~GHz) but they should be considered for broadband applications.


The TLS thermometers functionality and performance were first verified against a calibrated ruthenium oxide (RuOx) thermometer at temperatures down to 50 mK. A type-I TLS thermometer was adhered to a host chip with a 500~$\mu$m wide microstrip feedline (Fig. \ref{fig:design} B).
A type-II TLS thermometer was connected to a host chip with a CPW feedline by use of a single wire bond (Fig. \ref{fig:design} D).
An identical copy of the TLS thermometer resonator design was also directly fabricated on the bottom host chip. 
This allowed confirmation that the adhered TLS thermometers had the same response (i.e. the same temperature) as those that were embedded in the host chip, whose temperature was to be determined.  
We used a low-noise cryogenic amplifier at the 4~K stage of the cryostat to improve the signal-to-noise when reading out the resonators.  

The TLS thermometers agreement to Eqn \ref{eqn:TLS} and sensitivity were quantified by measuring fractional frequency shift (x~=~$\Delta f/f$), responsivity (R = dx/dT), and noise power ($S_{xx}$) from 50 mK to 2000 mK using a VNA and homodyne readout system \citep{Gao2007}.
Figure~\ref{fig:sensitivity} shows these measurements for an adhered 2~GHz type-II TLS thermometer.
s shown in Panel A of Fig. \ref{fig:sensitivity}, the temperature-dependent frequency shift data agrees very closely with Eqn.~\ref{eqn:TLS} for $T<1000$~mK, suggesting the theoretical TLS curve can be used as the standard calibration curve to high accuracy.
In fitting Eqn.~\ref{eqn:TLS} to the data, we have included both the measurement uncertainty in resonance frequency and the temperature uncertainty of the RuOx thermometer. 
From uncertainty analysis, we estimate the uncertainty of T reported by
this TLS thermometer to be $2\sigma_T \leq\pm$2 mK (50 mK <T< 250mK), $\leq\pm$6 mK (250 mK <T< 500mK) and $\leq\pm$14 mK (500 mK <T< 1000mK). 
These uncertainty values reflect the accuracy of the agreement between the TLS thermometer and the RuOx thermometer which we calibrate against. Any systematic bias in the RuOx will lead to a similar level of temperature bias in the TLS thermometer. In future work, we plan to calibrate the TLS thermometer against a primary Johnson noise thermometer to evaluate its absolute temperature accuracy.

Both the thermometers embedded in and adhered on top of the host chip showed strong agreement with Eqn. \ref{eqn:TLS}. A pair of type-I thermometers of the same design exhibited the same frequency shift and yielded nearly identical $F\delta_\mathrm{TLS}$ ($F\delta_\mathrm{TLS} =  $3.49$\times10^{-5}\pm1.7\times10^{-7}$ for the embedded and $F\delta_\mathrm{TLS}=$3.61$\times10^{-5}\pm2.1\times10^{-7}$ for the adhered), indicating that the adhered TLS thermometers are well thermalized to the host chip. 
Panel B to D in Fig. \ref{fig:sensitivity} show the responsivity, noise, and derived NET of the TLS thermometer. We see that both the responsivity and noise decrease with temperature, resulting in a relatively uniform NET = 50--70~$\mu$K/$\sqrt{\text{Hz}}$ over the temperature range 50--1200 mK.

To evaluate the long term stability, in a separate cooldown, five TLS thermometers were measured every minute for 48 hours with the bath temperature held at 150 mK.
The largest frequency deviation in all of the resonators translates to a $\pm$5~mK temperature variation.
Stability over even longer periods of time will be evaluated in the future, which is important for both TLS thermometer applications and the study of TLS physics\citep{burnett2019,Sarabi2016,Brehm2017}.

To examine the repeatability over multiple cryogenic cycles, four TLS thermometers were cooled down twice over a 1-week period (6-30 hours spent at atmosphere) for comparison.
We found that $f_0$ shifts down by 10s of kHz,
 which is comparable to the TLS induced frequency shift.
In contrast, $F\delta^0_{\mathrm{TLS}}$ is more stable, and varied by less than 3\%  (results in a 10 mK temperature difference occurring at the lowest temperatures).
Frequency shifts of a similar order have often been observed in the MKID community, which has yet to be systematically studied.
It is speculated that the negative frequency shift is associated with surface oxidation which leads to a reduction of metal thickness and an increase in the kinetic inductance\cite{Liu2017}. The surface oxidation has little effect on $F\delta^0_{TLS}$ for our devices because the TLS is dominated by the 25 nm amorphous SiO$_2$ on top of the Si substrate instead of the metal oxide. 
Because of the run-to-run resonance shift, a new calibration to pin down $f_0$ is required for each cooldown. 
However, a calibration to re-constrain $F\delta^0_{\mathrm{TLS}}$ between cooldowns is optional. 
In addition, intra-wafer non-uniformity\citep{McKenney2019} gives rise to a $\sim9.5\%$ variation in the fitted $F\delta^0_{\mathrm{TLS}}$ for resonators of identical design. Thus, for best performance, each resonator from the same wafer should be individually calibrated.

\begin{figure}[ht]
\includegraphics[trim=0.0in 0 0 0,clip,scale=.65]{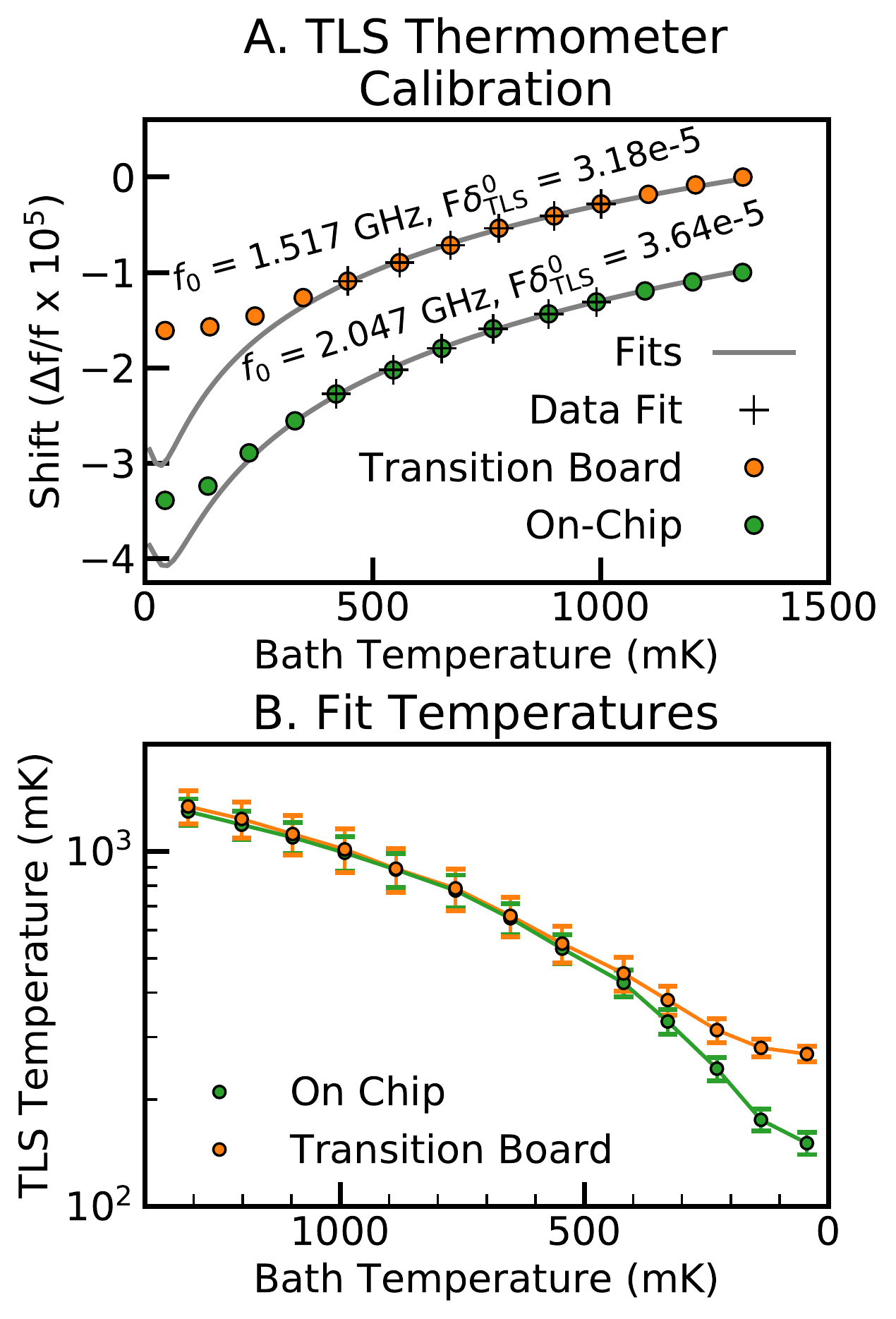}
\caption{\label{fig:calibration} TLS thermometer calibration in KIT experiment.  
Panel A: Resonant frequencies for both TLS thermometers were determined as a function of RuOx measured bath temperature during the dilution refrigerator cooldown.
Considering the potential lag between the TLS thermometer temperature and the bath temperature, as well as the cooldown rate, we have included an additional 10~mK temperature uncertainty in the calibration process.
Grey lines are fits to Eqn.~\ref{eqn:TLS} using data within the range 400~mK$<$T$<$1000~mK, and are used for subsequent temperature determination. 
Deviations from the fit at T$<$300~mK are attributed to true temperature differences from the RuOx due to excess heating.
Panel B: On-chip and transition board temperature as a function of RuOx bath temperature, showing temperature decoupling.
}
\end{figure}


\begin{figure}
\includegraphics[trim=0.0in 0 0 0,clip,scale=.69]{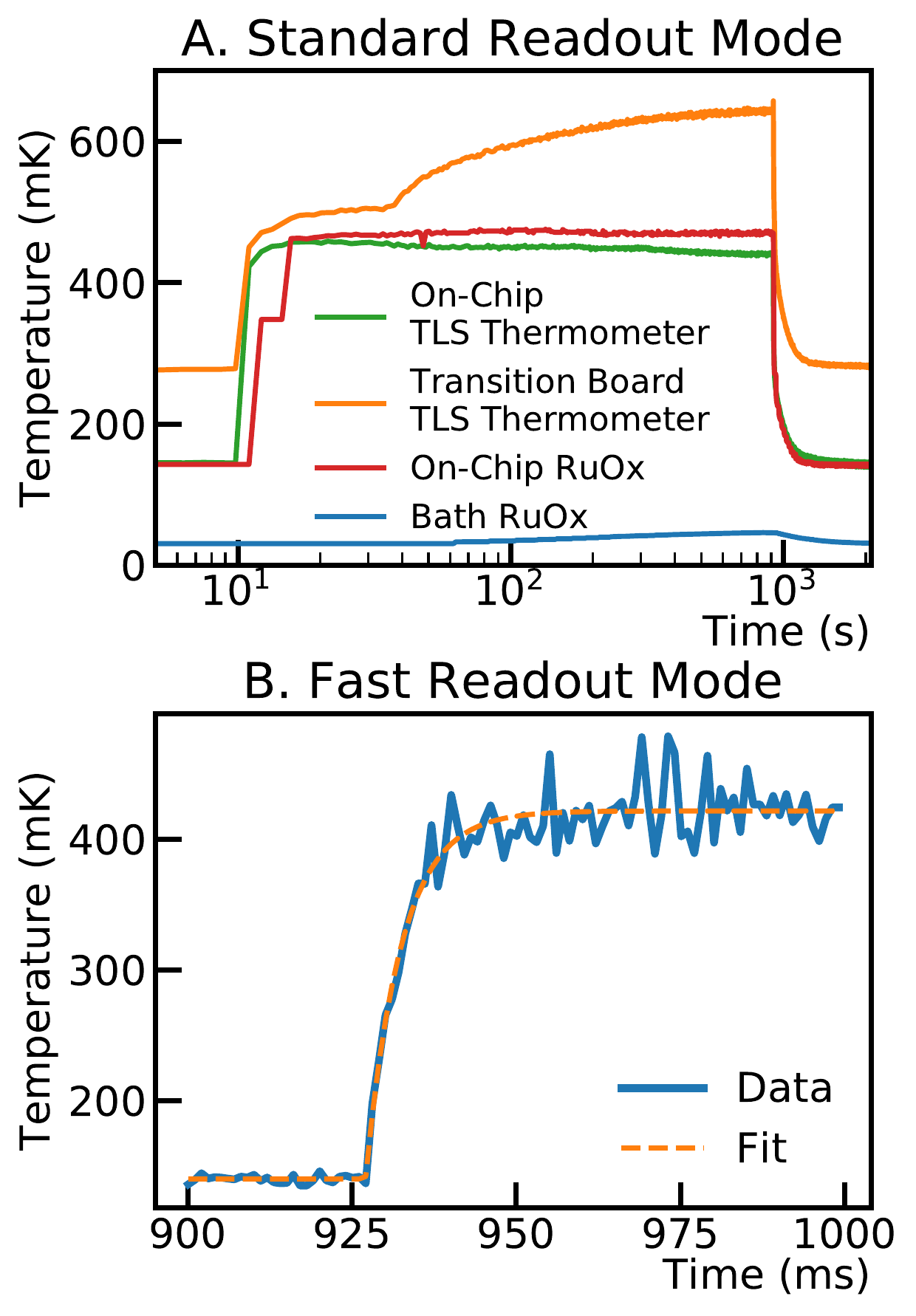}
\caption{\label{fig:slow} 
TLS thermometer measurements of KIT heating. 
Panel A shows slow readout (1.2~Hz) in which the KIT pump was turned on (off) at 10~s (900~s). 
The on-device thermometers record substantial heating whereas the stage thermometer monitoring the bath temperature (\textit{Bath RuOx}) shows only a 15 mK rise.  
The step in the \textit{On-Chip RuOx} curve at $\sim$~15 s is an artifact of the filtering used in the commercial readout.
Panel B shows the heating experiment in fast readout mode, sampled at 1~kHz. 
The KIT pump is turned on at approximately 927 ms, and the exponential temperature rise is captured by the TLS noise thermometer. Here we see a much smaller standard deviation of data for the pump-off state than the pump-on state. This is due to the sensitivity degrading effect associated with the probe tone going from on-resonance to off-resonance.
Tone tracking or a multiple tone read out would result in more uniform sensitivity.
}
\end{figure}

To demonstrate the utility of these thermometers, we used TLS thermometers to probe the thermal environment of a KIT amplifier package.  
KITs are active devices that utilize the nonlinearity of superconductors for parametric amplification \citep{Eom2012,Chaudhuri2015}.
A strong pump tone is required in the traditional four-wave mixing (4WM) KIT (10s of uW) \citep{bockstiegel2014development}.
In previous testing, when the pump tone is turned on, we found an increase in bath temperature. 
This indicated that heating was occurring but we could not quantify the temperature of the KIT. 
Accurate measurement of the actual chip temperature of KITs is desired because the chip temperature can play an important role in the achieved noise properties. 
To investigate the chip temperature, two TLS thermometers and a conventional RuOx resistive thermometer were attached to the KIT package (Fig.~\ref{fig:design}E). 
One TLS thermometer was attached to the KIT and one was adhered to the microstrip portion of the SMA-to-device transition circuit board. 
The KIT was mounted to the coldest stage of a dilution refrigerator with 1 dB attenuators on the input and output lines for thermalization at the 4~K stage. 
This configuration is suited to device gain characterization rather than system noise measurement. For this test no cryogenic amplifiers were used.   

The TLS thermometers in this experiment had not been previously calibrated.
As such, a calibration was performed in situ as the dilution refrigerator cooled (Fig. \ref{fig:calibration}). 
We find that, even without an active pump, the KIT fails to reach base temperature. 
This can be seen from the deviation of the data points from the TLS theoretical curves in Fig. 3A and the deviation from a straight line in Fig. 3B at the lowest temperatures.
Furthermore, the observation that the transition board is hotter than the device itself suggests that the heat source is due to thermal conduction along the coaxial lines, originating from components at higher temperature stages such as the 1dB attenuators.  

We performed two additional measurements to understand KIT heating from the rf pump, and the results are shown in Fig.~\ref{fig:slow}.  
For both measurements, a 6.5~GHz, -11.2~dBm (76~$\mu$W) rf pump was applied to the KIT, which yields $\sim$15dB of gain.   
In the first measurement, TLS thermometer temperatures were sampled every 1.2 s, while simultaneously monitoring both the bath and chip-adhered RuOx thermometers.   
Each TLS thermometer temperature measurement consisted of a 200 point VNA S$_{21}$ frequency sweep that was fit to extract the resonant frequency \citep{Gao2007}.  
Both the chip-adhered RuOx and the chip-adhered TLS thermometer showed rapid heating and cooling to and from $\sim$450~mK. 
Fifteen minutes after the activation of the pump, the RuOx and the TLS thermometer temperature differed by 30~mK. 
This temperature disparity may be real or due to a systematic effect in either thermometer. 
The transition board TLS thermometer showed quick heating to 500 mK followed by slower heating to above 600~mK, possibly from heat conduction along the coaxial lines from the heating of other microwave components at higher temperature stages. 

In the second measurement, we used the chip-adhered TLS thermometer in fast readout mode to resolve quick temperature changes (Fig.~\ref{fig:slow} B).
In this mode, the VNA is fixed at a single frequency $f_p$ and measures $S_{21}(f_p)$ at 1~kHz sampling rate, similar to the typical homodyne configuration used for MKID readout. 
We chose to use this unusual VNA operation mode to show that for simple measurements no additional instruments (synthesizer, amplifiers, or mixers) dedicated to the TLS thermometer are needed.
Because the KIT is a nonlinear transmission line, an overall phase change in S$_{21}$ occurs when the pump is turned on\citep{Eom2012,Vissers2016,Malnou2020}, This complicates resonator fitting in the complex S21
plane.
For this reason, we determine the frequency shift from the magnitude of transmission, $|S_{21}(f_p)|$, with the help of the calibration data, $|S_{21}(f_p;T)|$, measured during the cooldown.
As shown in Fig.~\ref{fig:slow} B, we find an exponential rise in the on-chip temperature with a fitted time constant $\tau$~=~5.5~ms when the pump is turned on. 
Because the resonator ring-down time is 1000 times faster, we determine that $\tau$ is a thermal time constant and likely associated with the KIT chip determined by the heat capacity of the KIT chip and thermal conduction of the KIT chip to the packaging.
As such, $\tau$ may be treated as an upper limit to the TLS thermometer thermal time constant. 

The three TLS thermometer on-chip measurements unambiguously quantify the heating of the KIT device, and furthermore indicate potential sources of heating to target for future mitigation efforts.  
These insights were made possible by TLS thermometers.
Now that the concept has been validated, the next iteration of KIT devices may include TLS thermometers co-fabricated within the KIT device.

While TLS thermometers will not replace conventional resistive or diode thermometers in sub-kelvin cryogenics, in certain situations they are more convenient.
In particular, the TLS thermometers provide a powerful tool below 1000~mK to measure the on-chip temperature of cryogenic microwave devices, such as microwave kinetic inductance detectors \citep{Zmuidzinas2012}, microwave SQUID multiplexers\citep{Mates2011}, and qubits \citep{Kjaergaard2020}. 
The thermometers can be frequency multiplexed alongside these devices utilizing the existing feedline and read out with the same infrastructure, as demonstrated with the 4WM KIT in this study.

The TLS thermometers are simple to design and implement. 
This application requires only adhesion to an exposed microstrip or a single wire bond to the feedline.
The miniature footprint allows the use in very small packages or even for co-fabrication with other superconducting microwave devices.   
Additionally, the absence of dc wires avoids potential sources of heating and electromagnetic interference.  
In many situations, temperature measurements with TLS thermometers will not require additional readout electronics.
Because the thermometers are read out in the frequency domain, many thermometers at different frequencies can be simultaneously measured.
This could allow mapping temperature gradients across a large wafer.

Going further, we will validate the long-term repeatability of the temperature measurements, investigate the systematic errors, and study their magnetic susceptibility.

Future designs will explore the deposition of additional amorphous dielectrics onto the resonators to increase the TLS participation, thereby increasing the responsivity and improving the NET.
The data that support the findings of this study are available from the corresponding author upon reasonable request.


\bibliography{tls_thermometer}

\end{document}